\documentstyle[aps,twocolumn,epsfig,amsfonts]{revtex}
\begin{document}

\title{Spinning up and down a Boltzmann gas}

\author{D. Gu\'ery-Odelin}
\address{Laboratoire Kastler Brossel, D\'epartement de Physique de l'Ecole Normale Sup\'erieure}
\address{24, Rue Lhomond, F-75231 Paris Cedex 05, France}

\date{\today}
 
\maketitle
 
\begin{abstract}
\noindent
Using the average method, we derive a closed set of linear equations that describes the spinning up of an harmonically trapped gas 
by a rotating anisotropy. We find explicit expressions for the time needed to transfer angular momentum as well as the decay
time induced by a static residual anisotropy. These different time scales are compared with the measured nucleation time and lifetime of vortices
by the ENS group \cite{VortexENS}. We find a good agreement that may emphasize the role played by the non-condensed component in those
experiments.
\end{abstract}
 

\section*{Introduction}

Superfluidity of a Bose-Einstein condensate is naturally investigated by its rotational
properties \cite{Baym,scissors}. So far, two experimental schemes have been successfully applied
to gaseous condensates in order to generate quantized vortices \cite{VortexJila,VortexENS}.
The first one uses phase engineering by means of laser beams whereas  the second 
one is the analog of the ``rotating bucket"
experiment, initially suggested by Stringari \cite{STRINGARI96,Sandro2}.

In the latter,  atoms are first confined 
in a static, axially symmetric Ioffe-Pritchard magnetic trap
upon which a non-axisymmetric attractive dipole potential is superimposed by means of a 
strirring laser beam. In this paper, we address the question of the behaviour of the non-condensed component
for this experiment. We investigate the transfer of
angular momentum to an ultracold harmonically confined gas by such a time dependent potential.
We also study possible mechanisms for its dissipation.
First (Sec. I), we recall the Lagragian and the Hamiltonian formalism for a single particle
in a rotating frame. Then in Sec. II, we give the 
expression for the rotating potential that has been used in our study. We briefly expose analytical results 
for the single particle trajectory in this potential in Sec. III. 
The rest of the paper deals with the crucial role played by collisions.
A classical gas that evolves in this 
potential thermalizes in the rotating frame, leading to a finite value of its mean angular momentum. 
We investigate this equilibrium state in Sec. IV.
In Sec. V, we derive the analytical expression for the time needed to spin up a classical gas with
an approach based on the average method \cite{AvM}. 
We evaluate the time needed  to transfer angular momentum and thus we deduce the characteristic time for vortex
nucleation {\it via} angular momentum transfer from the uncondensed to the condensed component.
Finally, in Sec. VI, we consider 
a related problem: what is the time needed
to dissipate a given angular momentum by a static residual anisotropy ?
We have in mind the role of the axial asymmetry of a magnetic trap (Ioffe-Pritchard or  time-orbiting potential  traps) induced by the presence of gravity.
We show that this effect may explain the finite lifetime of vortices.

\section{A Reminder on the rotating frame}

In this section, we recall the hamiltonian for a classical particle in a rotating frame 
characterized by the fixed rotation vector ${\bf \Omega}$.
Without loss of generality, we choose a rotation around the $z$ axis ${\bf \Omega}=\Omega {\bf e}_z$ and the same origin for
the rotating frame $\Re'$ as the one of the laboratory frame $\Re$ (see Fig. 1). 
In the following, quantities with a prime are
evaluated in the rotating frame. Coordinates in $\Re'$ are linked to coordinates in $\Re$ just
by a rotation of angle $\theta=\Omega t$:
\begin{equation}
\left(
\begin{array}{c}
x'  \\ y' 
\end{array}
\right)=
\left(
\begin{array}{cl}
 \;\;\;\cos \Omega t & \sin \Omega t  \\ -\sin \Omega t  &  \cos\Omega t 
\end{array}
\right) 
\left(
\begin{array}{c}
x  \\ y 
\end{array}
\right)
\nonumber
\end{equation}
The lagrangian for the single particle movement in the rotating frame reads \cite{Landau1}:
\begin{eqnarray}
{\cal L}'=\frac{1}{2}mv'^2+\frac{1}{2}m({\bf \Omega}\times {\bf r})^2 
+ m{\bf v}'.({\bf \Omega}\times {\bf r})
-V_{\rm ext},
\label{lagr}
\end{eqnarray}
where $V_{\rm ext}$ is the potential energy that describes the role of external forces.
The correspondance between the laboratory and rotating frames for momentum, angular momentum and the hamiltonian is given by:
\begin{eqnarray}
 {\bf p}' & = & \frac{\partial {\cal L}'}{\partial {\bf v}'}=m{\bf v}'+m{\bf \Omega} \times {\bf r}={\bf p} \nonumber\\
{\bf L}' & = & {\bf r} \times {\bf p'}={\bf r}\times {\bf p}={\bf L}    \nonumber\\
{\cal H}' & = & {\bf p}'.{\bf v}'-{\cal L}'={\cal H}-{\bf \Omega}.{\bf L}
\label{srot}
\end{eqnarray}
Note that the angular momentum as well as the momentum are the same in both frames, but
the link between the momentum and the velocity differs.
Using the hamiltonian formalism, one can easily extract the equation of motion in the rotating frame:
\begin{equation}
m\frac{d{\bf v}'}{dt}={\bf F}_{\rm ext}+{\bf F}_{\rm cen}+{\bf F}_{\rm cor}
\end{equation}
where ${\bf F}_{\rm ext}$ refers to the external force derived from the potential $V_{\rm ext}$,
${\bf F}_{\rm cen}$ is the centrifugal force:
\begin{equation}
{\bf F}_{\rm cen}=-m{\bf \Omega}\times ({\bf \Omega}\times {\bf r})=-
\mbox{\boldmath $\nabla$}V_{\rm cen}
\end{equation}
with $V_{\rm cen}=-m\Omega^2(x^2+y^2)/2$,
and ${\bf F}_{\rm cor}$ is the Coriolis force:
\begin{equation}
{\bf F}_{\rm cor}=-2m{\bf \Omega}\times {\bf v}'.
\end{equation}

Note that all formulas of the system (\ref{srot}) are still valid even for a time-dependent rotation
vector ${\bf \Omega}$. 

\section{The rotating trap}

In Bose-Einstein condensation experiments, the magnetic confinement is axially symmetric. In order to spin up
the system, one possibility consists of breaking this symmetry by superimposing a small rotating anisotropy
as initially suggested by S. Stringari \cite{STRINGARI96}.
 This breaking of the rotational invariance of the external
potential can be carried out experimentally by adding a
rotating stirring dipolar beam to the magnetic field of the trap \cite{VortexENS}.
This combination of light and magnetic trapping induces the 
following harmonic external potential (see Fig. 1) \cite{Oxford}:
\begin{equation}
V_{\rm ext}=\frac{m\omega_0^2}{2}\bigg((1+\epsilon)x'^2+(1-\epsilon)y'^2+\lambda^2
z^2\bigg)
\label{vext}
\end{equation}
where we have defined  the geometric parameter $\lambda=\omega_z/\omega_0$,
which is responsible for the shape of the cloud. For instance, if we set $\epsilon=0$,
the trap is isotropic for $\lambda=1$, cigar-shaped  for $\lambda \ll 1$, and disk-shaped 
for $\lambda \gg 1$. 
The potential (\ref{vext}) is time-dependent if expressed with laboratory coordinates $(x,y,z)$, but static
in the rotating frame, {\it i.e.} in terms of $(x',y',z)$.

\section{Single particle trajectory}

Let us first investigate the single particle trajectory.
Equations in the laboratory frame and for the transverse coordinates are given by:
\begin{equation}
\left(
\begin{array}{c}
\ddot{x}  \\ \ddot{y} 
\end{array}
\right)=-\omega_0^2 \left(
\begin{array}{c}
x  \\ y 
\end{array}
\right) -\epsilon\omega_0^2 
\left(
\begin{array}{cl}
\cos 2\Omega t & \;\;\;\sin 2\Omega t   \\ \sin 2\Omega t  & -\cos 2\Omega t 
\end{array}
\right) 
\left(
\begin{array}{c}
x  \\ y 
\end{array}
\right)
\nonumber
\end{equation}
In the rotating frame, the corresponding equations are time independent. They
can be rewritten by means of the complex quantity: $\xi'=x'+iy'$:
\begin{equation}
\ddot{\xi}'+2\Omega i\dot{\xi}'+(\omega_0^2-\Omega^2)\xi'+\epsilon
\omega_0^2\bar{\xi}'=0 
\label{eqz}
\end{equation}
where $\bar{\xi}$ denotes the complex conjugate of $\xi$.
The second term in Eq. (\ref{eqz}) accounts for the Coriolis force, the centrifugal force
leads to a reduction of the harmonic strength (third term), and
the last term is the contribution of the asymmetry.
The stability of the single particle movement is extracted from the
dispersion relation of (\ref{eqz}). We find a window of instability 
around the value $\Omega=\omega_0$, {\it i.e.} for 
$\Omega\in[\omega_0\sqrt{1-\epsilon}; \omega_0\sqrt{1+\epsilon}]$.
For $\Omega <\omega_0\sqrt{1-\epsilon}$, the stability is essentially 
ensured by the harmonic trapping even if reduced by the centrifugal force, whereas for $\Omega
>\omega_0\sqrt{1+\epsilon}$ the Coriolis force plays a crucial role in the
stabilization of the trajectory. The latter effect is similar to magnetron stabilization
in a Penning trap for ions \cite{magnetron}.

\section{Equilibrium state of the gas}

In practice, experiments are carried out in the so-called collisionless regime, 
since on average an atom undergoes less than a collision during a transverse oscillation period. 
However, collisions are of course essential to explain the dynamics of the gas induced by the potential
(\ref{vext}).
We consider the situation in which
a gas at a given temperature $T_0$ is initially at rest in the lab frame, and
at $t=0$ the axial asymmetry $\epsilon$ is spinned up at a constant angular
velocity ${\bf \Omega}$ as  explained above. If $\Omega
<\omega_0\sqrt{1-\epsilon}$  one expects that elastic collisions will ensure the
thermalization of the gas in the rotating frame. This equilibrium state is
defined since a minimum of the effective potential $V_{\rm eff}=V_{\rm ext}+V_{\rm cen}$ always exists
in this range of values for $\Omega$.
In Fig. 2, we compare the stability of a single particle (upper
part of the diagram) with that of the interacting gas.
In the rotating frame, one can compute equilibrium quantities by means of
the Gibbs distribution $\rho$ which reads \cite{Landau5}: 
\begin{equation}
\rho ({\bf r}',{\bf v'}) \propto e^{-{\cal H}'({\bf r}',{\bf v'})/k_BT_0}
\label{gd}
\end{equation}
where ${\cal H}'$ is given by:
\begin{equation}
{\cal H}'=\frac{mv^{\prime 2}}{2}+V_{\rm ext}(x',y',z)-\frac{m\Omega^2}{2}(x'^2
+y'^2)
\end{equation}
As regards statistical properties of the gas, the rotation is equivalent
to a reduction of the effective transverse frequencies of the trap due
to the contribution of the centrifugal force. The Coriolis force plays no role
for the equilibrium state. 
During the thermalization, the mean angular momentum per particle increases from
zero to its asymptotic value $\langle L_z\rangle$. This last quantity is 
easily derived from the Gibbs distribution (\ref{gd}):
\begin{equation}
\langle L_z\rangle=m\Omega\langle x^2+y^2 \rangle=
\frac{2k_BT_0\Omega(\omega_0^2-\Omega^2)}{(\omega_0^2-{\Omega}^2)^2
-\epsilon^2\omega_0^4}
\label{eqlz}
\end{equation}
For typical experimental parameters $\omega_0/2\pi=200$ Hz, $\Omega=\omega_0/2$,
$\epsilon=0.05$ and $T_0=1$ $\mu$K, the angular momentum per particle
$\langle L_z\rangle /\hbar\simeq 150 $ unlike the superfluid part for which for instance the
angular momentum is equal to $\hbar$ per particle in the presence of one
vortex. 

To check that the gas undergoes a full rotation, one can  search for a displacement of the critical temperature
induced by centrifugal forces \cite{Sandro2}, or  perform a time-of-flight measurement. In the latter case, one expects that the ratio between $x$
and $z$ size scales as $({1-\Omega^2/\omega_0^2})^{-1/2}$
for long time expansion.

\section{Time needed to reach equilibrium}

In this section we derive the expression for the time $t_{\rm up}$ needed to reach the
equilibrium state in the rotating coordinate system. 
In other words,
$t_{\rm up}$ corresponds to the time needed to build correlations between
$x$, $v_y$, $y$ and $v_x$.  To estimate this time, we use
the classical Boltzmann equation. Our analysis relies on the use of the average method as explained in
\cite{AvM}. For instance, the equation for $\langle x'v_{y'}-y'v_{x'}\rangle$ involves $\langle x'y'\rangle$
which itself is coupled to $\langle x'v_{y'}+y'v_{x'}\rangle$ and so on.
Terms that do not correspond to a conserved quantity in a binary elastic collision lead to
a non zero contribution of the collisional integral. For instance, a mean value such as
$\langle v_{x'}v_{y'}\rangle$ involves the occurence of quadrupole deformations in 
the velocity distribution which make the contribution $\langle v_{x'}v_{y'}I_{\rm coll}\rangle\neq 0$.
Here, $I_{\rm coll}$ stands for the collisional kernel of the Boltzmann equation \cite{Huang}.
At this stage, we perform a gaussian ansatz for the distribution function.
After linearization this quadrupolar contribution results in the
so-called relaxation time approximation \cite{Huang,Smith}:
$$
\langle v_{x'}v_{y'}I_{\rm coll}\rangle = -\frac{\langle v_{x'}v_{y'}\rangle}{\tau}
$$
This method leads to a closed set of $13\times 13$  linear equations (see
Appendix), and provides furthermore an explicit link between the relaxation time
$\tau$ and the collisional rate in the sample.

It is worth emphasizing that the dynamic transfer of angular momentum
to a classical gas by rotating a superimposed axial anisotropy involves a coupling
between all quadrupole modes. The average method is fruitful in the sense that it
yields  a closed set of equations when one deals with only quadratic moments, namely:
monopole mode, scissors mode, quadrupole modes, ...
Non-inertial forces are linear in position or velocity, and thus give rise only
to quadratic moments using the average method.

Although unimportant for equilibrium properties, the Coriolis force plays
a crucial role for reaching equilibrium. 
Fig. 3 depicts  a typical thermalization of the
gas in the rotating frame, leading to the equilibrium value
(\ref{eqlz}) of the angular momentum, obtained by a numerical integration of the 
$13\times 13$ set of equations. After a tedious but straightforward
expansion of the dispersion relation, one extracts the smallest eigenvalue 
that drives the relaxation for a given collisional regime. 
First we focus on the regime in which
experiments are performed: collisionless ($\omega_0\tau\simeq 10$) and with
$\epsilon^2 \ll \Omega^2/\omega_0^2$. Denoting
$t^{\rm CL}_{\rm up}$ as the time needed to spin up the gas in this regime, we
find: 
\begin{equation} 
t^{\rm CL}_{\rm up}= \frac{8\tau}{\epsilon^2}\left(
\frac{\Omega}{\omega_0}\right)^2
\end{equation}
The result is independent of the geometrical aspect ratio of the trap $\lambda$
as physically expected. Using the same numerical values as in the previous, we deduce 
that this time is very long $t^{\rm CL}_{\rm up}\simeq 15$ s.
Note that to transfer just $\hbar$ of angular momentum per particle, one needs only 100 ms.
This time is on the order of the nucleation time for vortices that has been experimentally observed by the ENS group \cite{nuclea}.
One should nevertheless be carefull since the non-condensed component
is actually a Bose gas rather than a classical one that evolves in a non harmonic potential because
of the mean field potential due to the condensed component.

In the hydrodynamic limit, the characteristic time for spinning up is: 
$$
t^{\rm HD}_{\rm up}=1/(2\epsilon^2\omega_0^2\tau).
$$
So far this regime is not accessible for ultracold atom experiments
since inelastic collisions prevent the formation of very high density samples.
We recover here the special feature of the hydrodynamic regime, {\it i.e.} the time needed to reach
equilibrium increases with the collisional rate.
In Fig. 4  we have reported the evolution
of $t_{\rm up}$ as a function of $\omega_0\tau$ from a numerical integration of
the $13\times 13$ system.
The smallest value is obtained between the collisionless and
hydrodynamic regimes, as is usual for the relaxation of a thermal gas \cite{AvM}.

\section{Time needed to dissipate a given angular momentum}

Consider a gas with a given angular momentum, obtained for instance as explained
before. If this gas evolves in an axially symmetric trap, the angular momentum
is a conserved quantity. In contrast, if a small asymmetry $\epsilon$
exists between the $x$ and $y$ spring constants, the angular momentum is no
longer a conserved quantity and it is thus
dissipated. We call $t_{\rm down}$ the typical time for the relaxation of 
the angular momentum. This problem is very different from the one we faced previously
since the rotating frame is not an inertial frame. We thus expect $t_{\rm down}\neq
t_{\rm up}$. As in the previous treatment, this problem can be computed
with the average method. The corresponding equations are nothing but the 
scissors mode equations \cite{scissors}, {\it i.e.} a linear set of 4 equations 
involving $\langle xy\rangle$, $\langle xv_y-yv_x\rangle$, $\langle xv_y+yv_x\rangle$
and $\langle v_xv_y\rangle$ (see Appendix). Searching a solution of this system of the form $\exp(-\lambda t)$, one finds
\begin{equation}
\lambda=\frac{1}{4\tau}\bigg(1-\sqrt{1-\epsilon^2/\epsilon_c^2}\bigg),
\label{tthdown}
\end{equation}
where the critical anisotropy is related to $\tau$ and $\omega_0$ by $\epsilon_c=1/(4\omega_0\tau)$.
In Fig. (5), we plot different curves for the relaxation of angular momentum depending
on the value of $\epsilon$ with respect to $\epsilon_c$. For $\epsilon<\epsilon_c$ (long dashed line), one has a purely
damped relaxation; in the limiting case $\epsilon \ll \epsilon_c$, $t_{\rm down}\simeq 1/(2\epsilon^2\omega_0^2\tau)$.
On the contrary for $\epsilon> \epsilon_c$ (solid line), one has a damped oscillating behaviour; in the limiting case 
$\epsilon \gg \epsilon_c$, $t_{\rm down}\simeq  4\tau$, and the oscillating frequency is equal to $\epsilon\omega_0$.
Moreover, as we use a linear analysis, this decay time does not depend on 
the specific initial value of the angular momentum.
For the experiment described in \cite{VortexENS}, the axial asymmetry induced
by gravity is on the order of 1\% and the collisional rate is such that $\omega_0\tau\simeq 10$. The corresponding decay
time of the angular momentum is then evaluated
from Eq. (\ref{tthdown}) to be $\sim$ 500 ms. This value matches surprisingly well the typical observed lifetime
of a vortex. Therefore, one possible interpretation may be that the thermal part acts as a reservoir of angular momentum and
thus ensures the stabilization of the vortex as long as this thermal part is itself rotating significantly. On the opposite
when the angular momentum of the thermal part is zero, the vortices disappear.

\section*{Conclusion}

In this paper we have investigated the dynamic of transfer and dissipation
of angular momentum for a classical gas and in all collisional regimes.
We derive very different time scales for both
processes. The short time deduced to dissipate angular momentum
suggests a high sensitivity of the experiment described in \cite{VortexENS}
to residual static anisotropy (see different time scales between Figs. (3) and (5)).
In practice, as underlined above one  cannot avoid residual fixed
anisotropy due for instance to gravity. The competition between a rotating and a static
anisotropy may explain why
no evidence for a full rotation of the classical gas was experimentally reported
in Ref \cite{VortexENS}.

\section*{acknoledgements}

 I acknowledge fruitful discussions with V. Bretin, F. Chevy, J. Dalibard, 
K. Madison, A. Recati, S. Stringari and F. Zambelli. I am grateful to Trento's BEC group
where part of this work was carried out.

\begin{figure}
\begin{center}
\epsfig{file=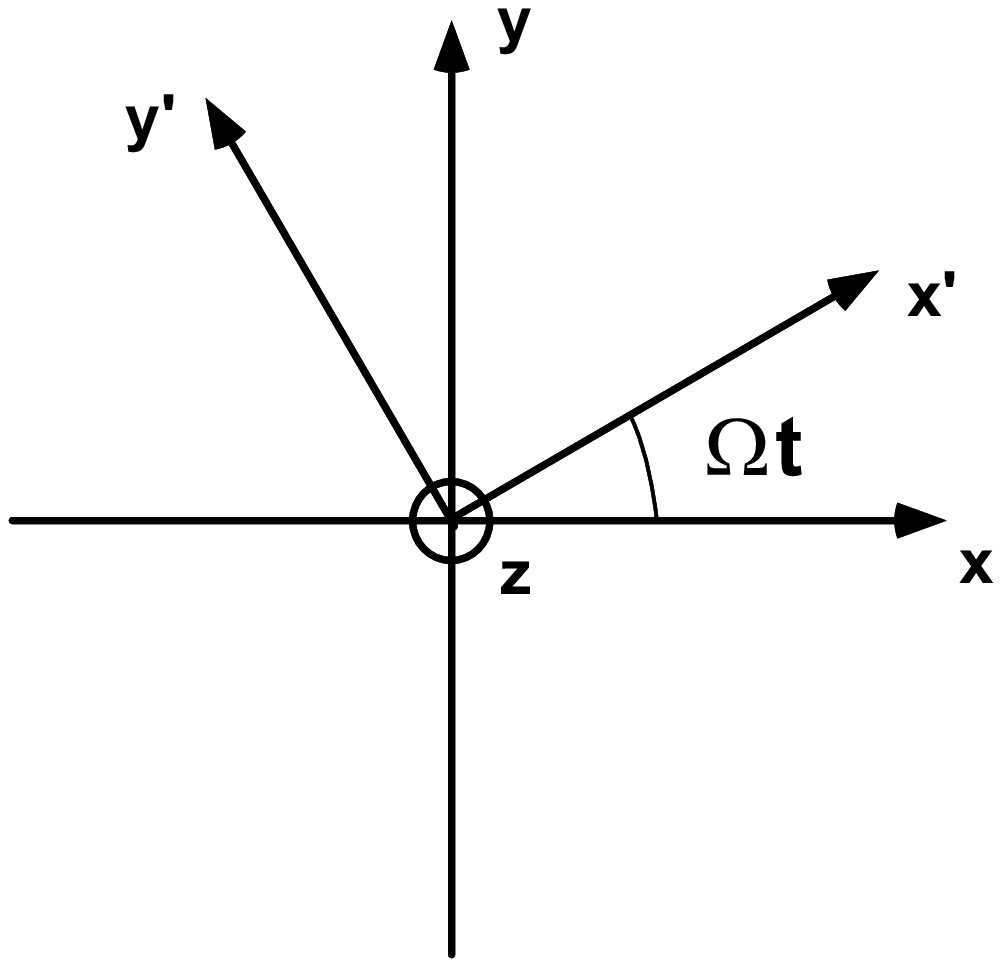, width=5cm}
\begin{caption}
{\sl Rotating frame with respect to laboratory frame.}
\end{caption}
\end{center}
\label{frame}
\end{figure}

\begin{figure}
\begin{center}
\epsfig{file=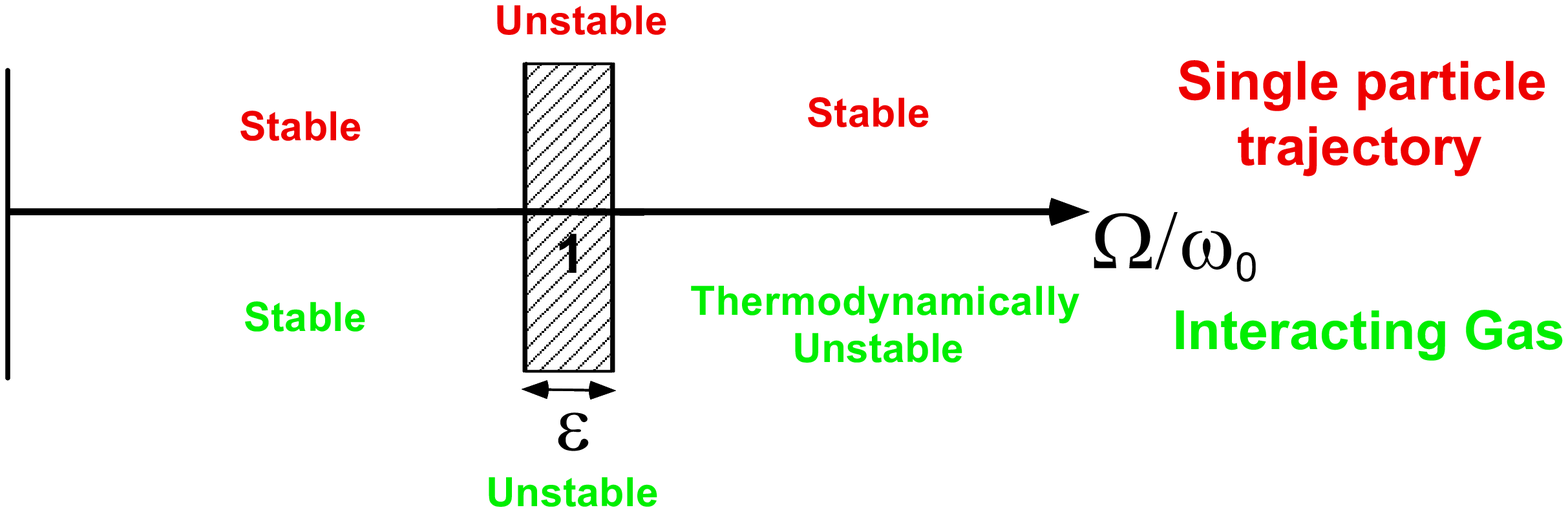, width=7cm}
\begin{caption}
{\sl Stability diagram for the single particle and for an interacting gas in the potential
with a rotating anisotropy.}
\end{caption}
\end{center}
\label{stab}
\end{figure}

\begin{figure}
\begin{center}
\epsfig{file=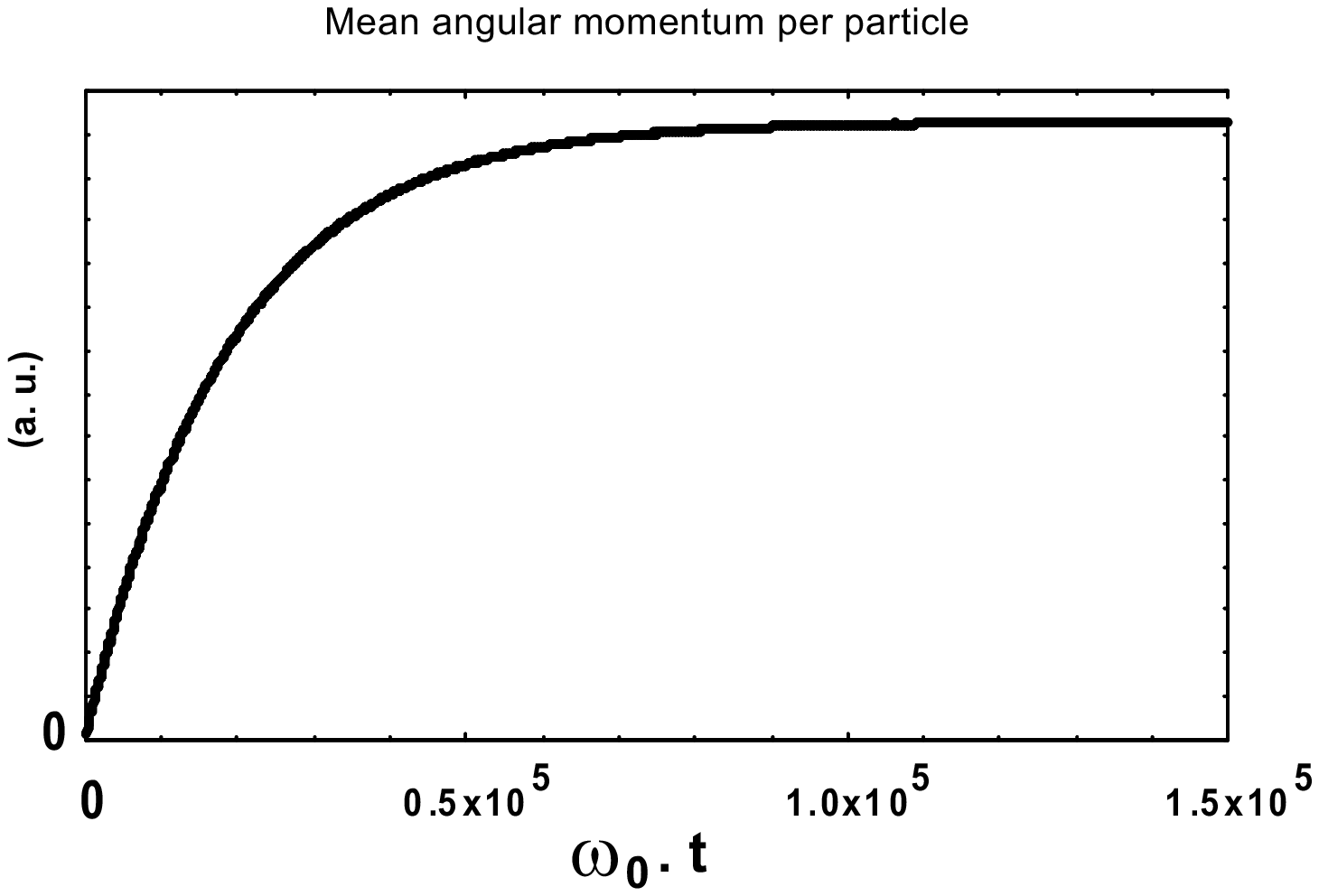, width=7cm}
\begin{caption}
{\sl Mean angular momentum per particle as a function of time.}
\end{caption}
\end{center}
\label{lup}
\end{figure}

\begin{figure}
\begin{center}
\epsfig{file=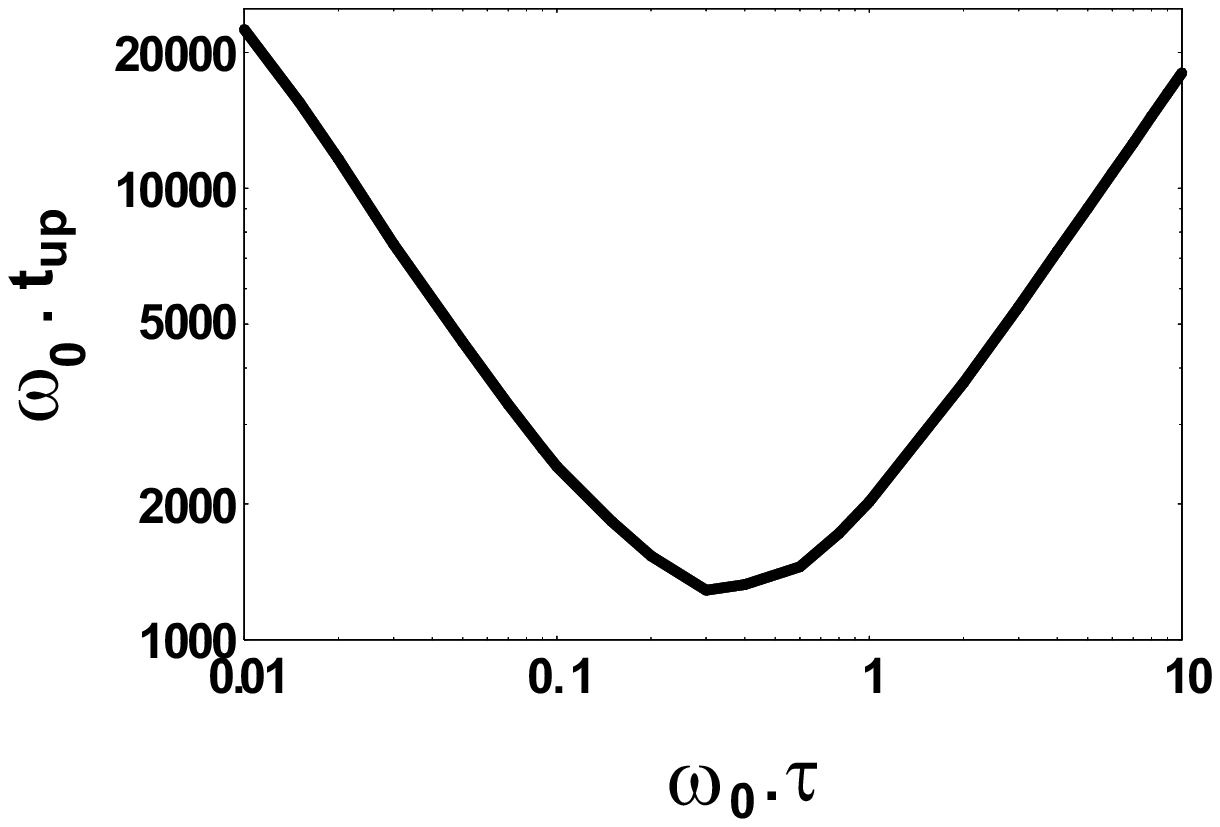, width=7cm}
\begin{caption}
{\sl Time needed to reach equilibrium by rotation of a small anisotropy. A minimum of $t_{\rm up}$ is
obtained in-between the collisionless and the hydrodynamic regimes.}
\end{caption}
\end{center}
\label{tup}
\end{figure}

\begin{figure}
\begin{center}
\epsfig{file=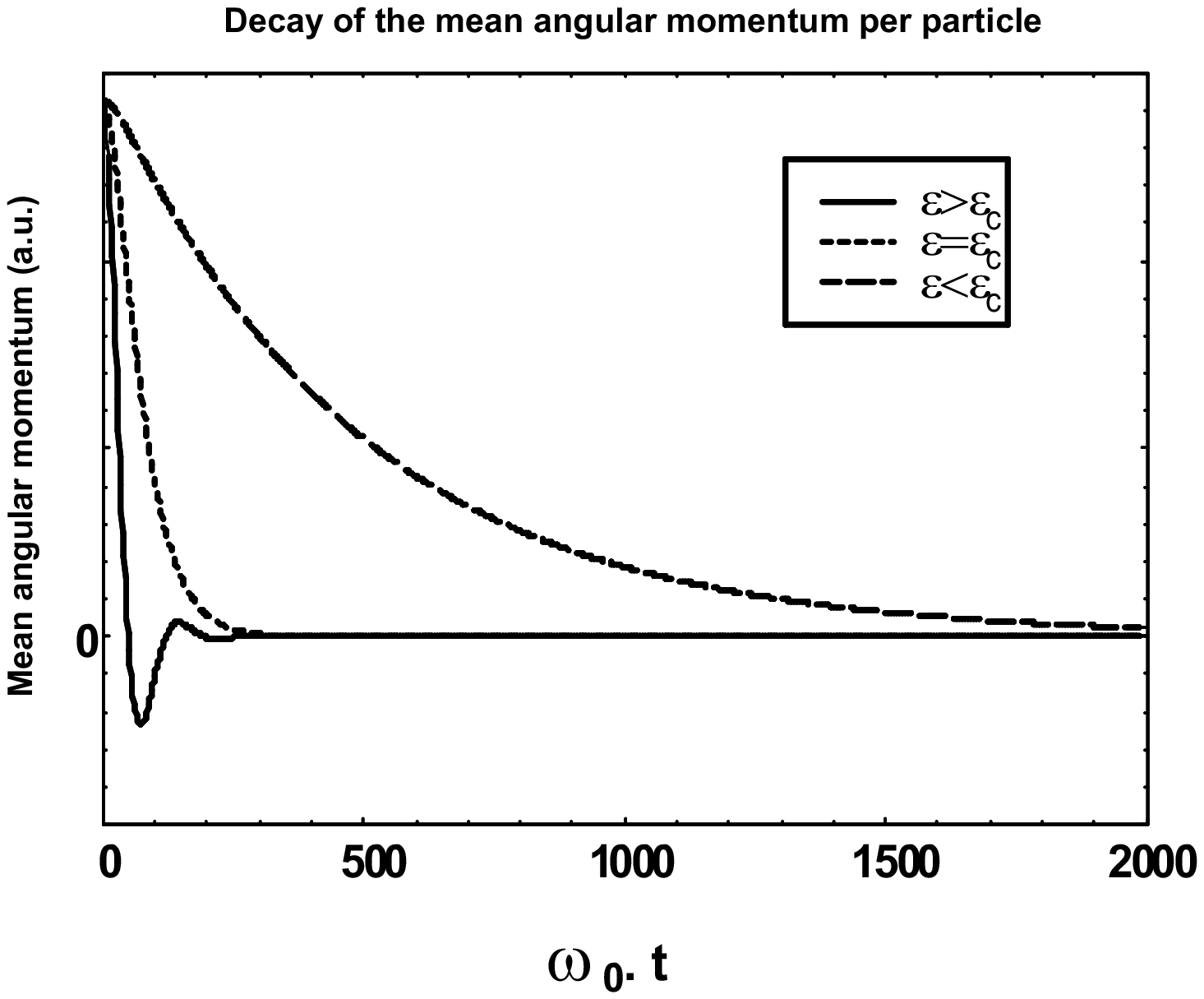, width=7cm}
\begin{caption}
{\sl Decay of angular momentum due to a small residual anisotropy. We distinguish three different regimes:
$\epsilon>\epsilon_c$ damped oscillations (solid line), $\epsilon=\epsilon_c$  (small dashed line)
the frontier with the region
of purely damped decay ($\epsilon<\epsilon_c$, long dashed line).}
\label{ldown}
\end{caption}
\end{center}
\label{ldown}
\end{figure}

\appendix
\section{APPENDIX}
\label{A}

Hereafter, the 13$\times$13 closed set of equations that describes the dynamic of
a classical gas induced by the rotating potential (\ref{vext}):
\begin{center}
\begin{eqnarray}
\frac{d\langle x'y'\rangle}{dt}-\langle x'v'_{y'}+y'v'_{x'}\rangle=0
\\
\frac{d\langle x'^2-y'^2\rangle}{dt}-2\langle x'v'_{x'}-y'v'_{y'}\rangle=0
\\
\frac{d\langle x'^2+y'^2+z'^2\rangle}{dt}-2\langle x'v_{x'}+y'v_{y'}+z'v_{z'}\rangle=0
\\
\frac{d\langle x'^2+y'^2-2z'^2\rangle}{dt}-2\langle x'v_{x'}+y'v_{y'}-2z'v_{z'}\rangle=0
\\
\frac{d\langle x'v_{y'}-y'v_{x'}\rangle}{dt}-2\omega_0^2\epsilon \langle x'y'\rangle
\nonumber\\
+\frac{4\Omega}{3}\langle x'v_{x'}+y'v_{y'}+z'v_{z'}\rangle
\nonumber\\
+\frac{2\Omega}{3}\langle x'v_{x'}+y'v_{y'}-2z'v_{z'}\rangle=0
\\
\frac{d\langle x'v_{y'}+y'v_{x'}\rangle}{dt}+2(\omega_0^2-\Omega^2)\langle x'y'\rangle
\nonumber\\
+2\Omega \langle x'v_{x'}-y'v_{y'}\rangle-2\langle v_{x'}v_{y'}\rangle=0
\\
\frac{d\langle x'v_{x'}-y'v_{y'}\rangle}{dt}+(\omega_0^2-\Omega^2)\langle x'^2-y'^2\rangle
\nonumber\\
+\frac{2\omega_0^2\epsilon}{3}\langle x'^2+y'^2+z'^2\rangle-2\Omega \langle x'v_{y'}+y'v_{x'}\rangle
\nonumber\\
+\frac{\omega_0^2\epsilon}{3}\langle x'^2+y'^2-2z'^2\rangle-\langle v_{x'}^2-v_{y'}^2\rangle=0
\\
\frac{d\langle x'v_{x'}+y'v_{y'}+z'v_{z'}\rangle}{dt}+\epsilon\omega_0^2 \langle x'^2-y'^2\rangle
\nonumber\\
+\frac{2-2\Omega^2+\lambda^2}{3}\omega_0^2\langle x'^2+y'^2+z'^2\rangle
\nonumber\\
+\frac{1-\Omega^2-\lambda^2}{3}\omega_0^2\langle x'^2+y'^2-2z'^2\rangle
\nonumber\\
-\langle v_{x'}^2+v_{y'}^2+v_{z'}^2\rangle-2\Omega \langle x'v_{y'}-y'v_{x'}\rangle=0
\\
\frac{d\langle x'v_{x'}+y'v_{y'}-2z'v_{z'}\rangle}{dt}+\epsilon\omega_0^2 \langle x'^2-y'^2\rangle
\nonumber\\
+\frac{2\omega_0^2}{3}(1-\lambda^2-\Omega^2)\langle x'^2+y'^2+z'^2\rangle
\nonumber\\
+\frac{\omega_0^2}{3}(1+2\lambda^2-\Omega^2)\langle x'^2+y'^2-2z'^2\rangle
\nonumber\\
-\langle v_{x'}^2+v_{y'}^2-2v_{z'}^2\rangle-2\Omega \langle x'v_{y'}-y'v_{x'}\rangle=0
\\
\frac{d\langle v_{x'}v_{y'}\rangle}{dt}+\epsilon\omega_0^2 \langle x'v_{y'}-y'v_{x'}\rangle+2\Omega \langle v_{x'}^2-v_{y'}^2\rangle
\nonumber\\
+(\omega_0^2-\Omega^2)\langle x'v_{y'}+y'v_{x'}\rangle=-\frac{\langle v_{x'}v_{y'}\rangle}{\tau}
\\
\frac{d\langle v_{x'}^2-v_{y'}^2\rangle}{dt}+2(\omega_0^2-\Omega^2)\langle x'v_{x'}-y'v_{y'}\rangle
\nonumber\\
-8\Omega \langle v_{x'}v_{y'}\rangle+\frac{4\epsilon\omega_0^2}{3}\langle x'v_{x'}+y'v_{y'}+z'v_{z'}\rangle+
\nonumber\\
\frac{2\epsilon\omega_0^2}{3}\langle x'v_{x'}+y'v_{y'}-2z'v_{z'}\rangle
=-\frac{\langle v_{x'}^2-v_{y'}^2\rangle}{\tau}
\\
\frac{d\langle v_{x'}^2+v_{y'}^2+v_{z'}^2\rangle}{dt}+2\epsilon\omega_0^2 \langle x'v_{x'}-y'v_{y'}\rangle
\nonumber\\
+\frac{2\omega_0^2}{3}(2+\lambda^2-2\Omega^2)\langle x'v_{x'}+y'v_{y'}+z'v_{z'}\rangle+
\nonumber\\
\frac{2\omega_0^2}{3}(1-\lambda^2-\Omega^2)\langle x'v_{x'}+y'v_{y'}-2z'v_{z'}\rangle=0
\\
\frac{d\langle v_{x'}^2+v_{y'}^2-2v_{z'}^2\rangle}{dt}+2\epsilon\omega_0^2 \langle x'v_{x'}-y'v_{y'}\rangle
\nonumber\\
+\frac{4\omega_0^2}{3}(1-\lambda^2-\Omega^2)\langle x'v_{x'}+y'v_{y'}+z'v_{z'}\rangle
\nonumber\\
+\frac{2\omega_0^2}{3}(1+2\lambda^2-\Omega^2)\langle x'v_{x'}+y'v_{y'}-2z'v_{z'}\rangle
\nonumber\\
=-\frac{\langle v_{x'}^2+v_{y'}^2-2v_{z'}^2\rangle}{\tau}
\end{eqnarray}
\end{center}
One can show from the gaussian ansatz that the relaxation time $\tau$ is the same
for all quadrupolar contributions.
In this system, linear terms in  $\Omega$ account for the Coriolis
force whereas quadratic terms in $\Omega$ are due to centrifugal force 
contributions. All quadrupolar modes are involved in this system.
In order to enlight the physics of this system, let us consider some limiting cases.
One can check the conservation of energy in the rotating frame $d\langle{\cal H}'\rangle/dt=0$.
The stationary state, obtained by setting all time derivatives of moments to zero,
is nothing but the equipartition law (see \cite{Landau5}, \S 44):
$(1-\Omega^2+\epsilon)\langle x'^2\rangle=(1-\Omega^2-\epsilon)\langle y'^2\rangle=\lambda^2\langle z^2\rangle
=\langle v_{x'}^2\rangle=\langle v_{y'}^2\rangle=\langle v_{z'}^2\rangle$.
If $\epsilon=0$ and $\Omega=0$, the trap is axially symmetric and one recovers the conservation of
the angular momentum. In addition the system gives rise to 3 independent linear systems on quadrupolar quantities: one
for the $m=2,\langle xy\rangle$ mode,
another one for  $m=2,\langle x^2-y^2\rangle$, and finally the $m=0$ mode that describes the coupling between
monopole ($\langle r^2\rangle$) and quadrupole mode ( $m=0,\langle x^2+y^2-2z^2\rangle$) \cite{AvM}.
Finally, one can recover the scissors mode equations for a classical gas by considering the case $\Omega=0$ and
$\epsilon\neq 0$ \cite{scissors}: eqs A1, A5, A6, A10.

\end{document}